\documentclass[10pt]{article}
\newcommand{\bsigma}{\mbox{\boldmath $\sigma$}}
\newcommand{\btau}{\mbox{\boldmath $\tau$}}
\usepackage{graphicx}
\begin{document}
\begin{center}
{\Large \bf RECENT RESULTS IN CBF THEORY \\ FOR MEDIUM-HEAVY NUCLEI} \\
\vskip 1.0 cm
{C. BISCONTI~$^{1}$, G. CO'~$^{1}$, 
F. ARIAS DE SAAVEDRA~$^{2}$, 
A. FABROCINI~$^{3}$       
}
\vskip 0.2 cm
{$^{1}$Dipartimento di Fisica, Universit\`a di Lecce \\
  and, \\
  Istituto Nazionale di Fisica Nucleare, sez. di Lecce\\
  Lecce, Italy\\
  $^{2}$Departamento de Fisica Moderna, Universidad de Granada, Spain\\
  $^{3}$Dipartimento di Fisica, Universit\`a di Pisa \\
  and, \\
  Istituto Nazionale di Fisica Nucleare, sez. di Pisa\\
  Pisa, Italy} 
\end{center}
\vskip 1.0 cm
\centerline{ABSTRACT}
{\small We extend the correlated basis
  functions theory (CBF) for nuclei with $N\neq Z$ and j-j coupling
  scheme. By
  means of the Fermi hypernetted chain integral equations, in
  conjunction with the single operator chain approximation (FHNC/SOC),
  we evaluate the ground state and the one-body densities for
  $^{40}Ca$,$^{48}Ca$ and $^{208}Pb$ nuclei. The realistic Argonne V8'
  two-nucleon potentials has been used. We compare the ground-state
  properties of these nuclei calculated  by using
  correlation functions with 
  ($f_{6}$ model) and without ($f_{4}$ model) tensor components.}
\vskip 1.5 cm
\noindent
The aim of this contribution is to report on the progresses
done in the framework of a project~\cite{Co92}-\cite{Fab00} 
aimed to apply the CBF theory to finite nuclear systems.
The starting point of the theory is the variational principle,
\begin{equation}
\delta E[\Psi]=
\delta \left[ \frac{<\Psi|H|\Psi>}{<\Psi|\Psi>} \right] =0
\label{eq:e}
\end{equation}
with the following ansatz for the expression of 
the many-body wave functions $\Psi$: 
\begin{equation}
\Psi(1,2,...,A)= G(1,2,...A)\,\Phi(1,2,...,A) 
\end{equation}
The search of the minumum of the energy functional is done by making
variations of the many-body correlation operator $G(1,2,...,A)$ and of
the single particle wave functions forming the Slater determinant
$\Phi(1,2,...,A)$.  The many-body correlation operator is supposed to
be described by a product of two-body operators $F_{ij}$:
\begin{equation}
G(1,2...,A)= S \prod_{i<j}F_{ij}
\end{equation}
where the operator $S$ symmetrizes the product.
The two-body correlations $F_{ij}$ have an operatorial dependence
analogous to that of the modern nucleon-nucleon interactions. We
consider $F_{ij}$ of the form:
\begin{equation}
F_{ij}=\sum_{p=1,8}f^{p}(r_{ij})\,O^{p}_{ij}
\label{eq:fij}
\end{equation}
where the involved operators are:
\begin{equation}
O^{p=1,8}_{ij}=
[1,\bsigma_{i}\cdot\bsigma_{j},S_{ij},({\bf L}\cdot {\bf S})]
\otimes
[1,\btau_{i}\cdot\btau_{j}]
\label{eq:op}
\end{equation}
In the above equation $\bsigma$ and $\btau$ indicate the usual spin
and isospin Pauli matrices, and $S_{ij}$ the tensor operator. 

The evaluation of the many-variables integrals necessary to calculate
the energy functional (\ref{eq:e}) is done by using the integral
summation technique known as Fermi hypernetted chain
(FHNC)~\cite{Ros82}, originally developed for infinite systems.  The
FHNC equations allows the sum of a set of infinite classes of
Mayer-like diagrams resulting from the cluster expansion of Eq.
(\ref{eq:e}).  The use of state dependent correlations, as those of
Eq. (\ref{eq:fij}), requires special attention because the various
correlation operators do not commute.  In our calculations this
difficulty is handled by considering, in addition to all the scalar
correlations, only those diagrams having chains of correlation
functions containing a single operator with $p > 1$.  This is the
so-called single operator chain (SOC) approximation~\cite{Pan79}.
\begin{table}[ht]
\begin{center}
\begin{tabular}{ccccccc}
\hline
& $F_{4}$ & $F_{6}$& $F_{4}$ & $F_{6}$& $F_{4}$ & $F_{6}$
\\
\hline
          & $^{40}$Ca                        & $^{40}$Ca              &
            $^{48}$Ca                        & $^{48}$Ca              &   
            $^{208}$Pb                       & $^{208}$ Pb           \\\hline
$T/A$     &38.61                          & 38.18                      &
           36.36                          & 36.21                      &    
           36.93                          & 36.21                       \\
$T_{jj}/A$  &  0.008                      & 0.006                      &
            0.16                          & 0.11                       &
            0.15                          & 0.09                        \\
$V_{8}/A$   & -51.81                      & -47.11                     &     
              -49.29                      & -45.29                     & 
              -54.17                      & -47.63                     \\
$V_{8jj}/A$ &  -0.0002                    & -0.0003                    &
               0.01                       & 0.11                       &
               0.02                       & 0.02                      \\
$V_{C}/Z$   & 3.78                        &  3.92                      &  
              2.55                        & 2.67                       & 
              9.58                        & 10.45                      \\
$V_{Cjj}/Z$ & 0.018                       &-0.0012                     &
             -0.016                       &0.008                       & 
             -0.01                        &-0.01                       \\ 
$E_{ee}/A$  &   0.47  & 0.46 &  0.45  & 0.44  &   0.34 & 0.32  \\ 
\hline
$E/A$       &-11.56                       & -6.73                      &   
             -10.99                       & -7.15                      &
             -12.97                       & -6.87                    \\
$E_{exp}$ &                               & -8.55                      &  
                                          &-8.66                       & 
                                          & -7.80                    \\ \hline
\end{tabular}
\caption {Contributions to the binding energies per nucleon  
   for the three nuclei
  considered. All the quantites are expressed in MeV.}
\label{tab:energies}
\end{center}
\end{table}

In the past, we have applied this theory to describe the ground state
of doubly closed shell nuclei by using two-body potentials and
correlations containing operator terms up to the tensor
components~\cite{Co92,Co94,Ari96,Fab98}, and also a three-body
interaction~\cite{Fab00}. These calculations have been limited to the
$^{16}$O and $^{40}$Ca nuclei since the formalism was developed in
$ls$ coupling scheme and for single particle wave functions equal for
both protons and neutrons.  Formally this situation is very similar to
that of the symmetric nuclear matter.

In this report we present the first results obtained by extending the
FHNC/SOC scheme for nuclei with different wave functions for protons
and neutrons and in $jj$ coupling representation.  The results of
these calculations have been done by using the realistic $v_{8}'$
Argonne two-body potential, but without any three-body force, which,
in any case, it will be soon implemented.  For this reason, the
results we show here, should not be considered fully realistic.  In
any case, already at this stage, they give interesting informations on
the nuclear structure.

The separated treatment of protons and neutrons requires that the
correlations should be considered separately in their spin and isospin
part:
\begin{equation}
F_{ij}=\sum_{k=1}^{3}\sum_{l=0}^{1}f_{2k-1+l}(r_{ij})O^{2k-1+l}_{ij}=
\sum_{l=0}^{1}(\vec{\tau}_{i}\cdot\vec{\tau}_{j})^{l}
       \sum_{k=1}^{3}f_{2k-1+l}(r_{ij})P^{k}_{ij}
\end{equation}
Another, in principle more obvious, consequence, is that the
FHNC/SOC equations should be almost triplicated to consider one- and
two-body densities of proton-proton, neutron-neutron, and isospin mixed
type. In $jj$ coupling these equations should be further extended to 
distinguish between two different types of statistical 
correlations, with parallel or antiparallel spin.

As already stated, the calculations have been done by using the
$v_{8}'$ reduction of the Argonne $v_{18}$ two-body
potential~\cite{Wir95}.  The operator structure of this potential is
given in Eq. (\ref{eq:op}).  The channels up to $p=6$ have been
treated within the FHNC/SOC formalism without any approximation, while
the contributions of the two spin-orbit channels have been calculated
by using perturbation theory~\cite{Fab00,Pan79}.

In the correlations the two spin-orbit channels have been neglected.
The results have been obtained by using a fixed set of single particle
wave functions and searching for the minimum by modifying the
correlations. The two-body correlation functions, have been generated
by solving a set of coupled Euler-Lagrange equations. The search of
the minimum has been done by changing only two parameters: the healing
distance $d_t$ of the two tensor channels, and another healing
distance $d_c$ related to all the other channels.  These are the
nucleonic relative distances where the various two-body correlation
functions reach their asymptotic values, one for $p=1$ and zero for
all the other cases.

We show in Tab. \ref{tab:energies} the results obtained for the
$^{40}$Ca, $^{48}$Ca and $^{208}$Pb nuclei. The single particle wave
functions have been generated by using a Woods-Saxon potentials whose
parameters have been taken from the literature~\cite{Ari96}.  For each
nucleus, in addition to the full calculation, whose results are
indicated in the $F_{6}$ columns, we also made calculations by using
only the first four central channels of the correlations. These are
the results shown by the $F_{4}$ columns.  The rows of the table
indicate the contributions of the various terms to the binding energy
per nucleon: $T$ is the kinetic energy, $V_8$ the nuclear interaction
term, $V_C$ the Coulomb term (here divided by the number of protons),
and finally $E_{ee}$ is the contribution of a relevant elementary
diagram that should be added to the FHNC/SOC calculations~\cite{Co92}
to properly satisfy the various sum rules.  All the terms with the
$jj$ labels indicate the contribution due to the antiparallel spin
densities.

A first remark, is that the $f_{4}$ calculations produce more binding
than the $F_{6}$ ones. The tensor terms of the correlation reduce the
binding as it is shown by the $V_8$ row. The kinetic energies have
similar values in all the calculations, and also the Coulomb terms are
not affected by the tensor correlation.
\begin{figure}[h]
\begin{center}
\includegraphics[scale=1.2]{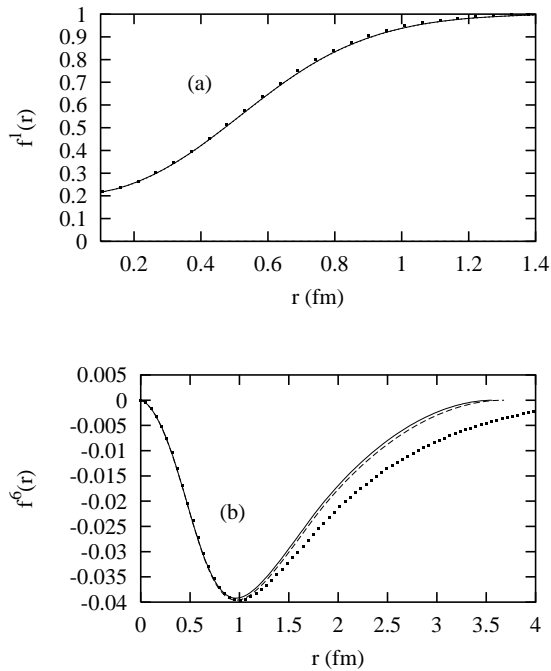}
\caption{The $f^{1}$,$f^{6}$ correlation functions. 
The $^{40}$Ca,$^{48}$Ca,$^{208}$Pb correlations are indicated
with solid , dashed  and dotted lines respectively. 
\label{fig:cor}}
\end{center}
\end{figure}

In Fig. \ref{fig:cor} we compare the correlation functions of the
three nuclei considered.  In the upper panel the scalar correlation
functions, are shown, while in the other panel the tensor-isospin
correlation functions, $p=6$ channel in Eq.(\ref{eq:op}), are
compared.  The scalar correlation functions are rather similar for all
the three nuclei considered, they heal at 1.4 fm. The tensor-isospin
correlations heal at larger distances and show a strong dependence on
the nucleus. The healing for the $^{208}$Pb is larger than that for
the other two nuclei.

\begin{figure}[h]
\begin{center}
\includegraphics[scale=1.2]{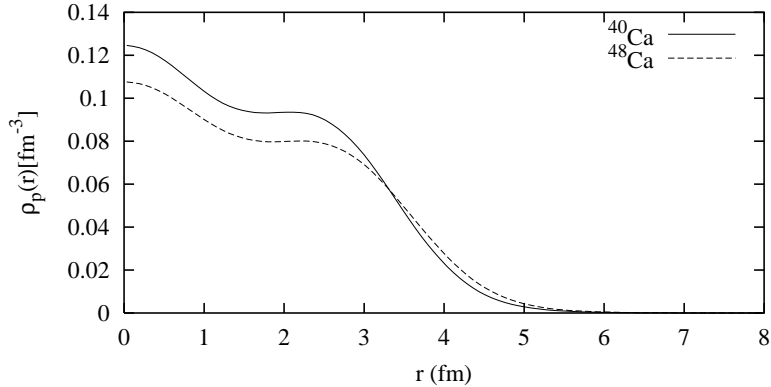}
\caption{Proton densities of the $^{40}$Ca and $^{48}$Ca nuclei.
\label{fig:dens}}
\end{center}
\end{figure}

Another interesting remark comes from observing the results
obtained by the same type of calculation in the various nuclei.
Kinetic energies and nuclear energies clearly show saturation properties
while the  repulsive Coulomb contribution increases with increasing
proton number. The comparison between the Coulomb terms of the two
calcium isotopes show a relatively large difference. This is 
surprising, but it demonstrates that the presence of a different number of
neutrons modifies the proton densities, as it is shown in Fig. 
\ref{fig:dens}.

A further remark about the results shown in Tab. \ref{tab:energies}
regards the contribution of the antiparallel spin terms, the $jj$
contributions. In general, these contributions are rather small and,
as expected, they become even smaller in $^{40}$Ca where all the
spin-orbit partners levels are saturated.

To conclude, we may say that the extension of the FHNC/SOC formalism
to treat nuclei with N$\neq$Z in $jj$ coupling scheme has been
successful. This has allowed us to perform calculations of binding
energies and density distributions of medium-heavy nuclei with neutron
exess such as $^{48}$Ca and $^{208}$Pb, by using a realistic two-body
potential.  To the best of our knowledge this is the first
microscopic calculation for the $^{208}$Pb nucleus. As already
mentioned, a fully realistic calculation should include also a
three-body interaction. The work in this direction has quite advanced
and we plan to obtain results in short time.


\end{document}